\documentclass{emulateapj}
\usepackage{amssymb}
\usepackage{apjfonts}
\usepackage{graphicx}
\usepackage{natbib}
\begin{document}
\title{Catalog of narrow $\rm C~IV$ absorption lines in BOSS (II): for quasars with $\rm z_{em} > 2.4$}
\shorttitle{Catalog of narrow $\rm C~IV$ absorption lines}
%\slugcomment{Draft Version}
\shortauthors{Chen et al.}

\author{Zhi-Fu Chen\altaffilmark{1, 2, 3}, Yi-Ping Qin\altaffilmark{1, 3, 4}, Ming Qin\altaffilmark{1}, Cai-Juan Pan\altaffilmark{1}, Da-Sheng Pan\altaffilmark{1}}

\altaffiltext{1}{Department of Physics and Telecommunication Engineering of Baise University, Baise 533000 China;
zhichenfu@126.com}

\altaffiltext{2}{Department of Astronomy, Nanjing University, Nanjing 210093, China}

\altaffiltext{3}{Center for Astrophysics,Guangzhou University, Guangzhou 510006, China}

\altaffiltext{4}{Physics Department, Guangxi University, Nanning 530004, China}

\begin{abstract}
As the second work in a series of papers aiming to detect absorption systems in the quasar spectra of the Baryon Oscillation Spectroscopic Survey, we continue the analysis of Paper I by expanding the quasar sample to those quasars with $z_{\rm em}>2.4$. This yields a sample of 21,963 appropriate quasars to search for narrow $\rm C~IV\lambda\lambda1548,1551$ absorptions with $W_r\ge0.2$ \AA~ for both lines. There are 9708 quasars with at least one appropriate absorption system imprinted on their spectra. From these spectra, we detect 13,919 narrow Civabsorption systems whose absorption redshifts cover a range of $z_{\rm abs}=1.8784$ --- $4.3704$. In this paper and Paper 1, we have selected 37,241 appropriate quasars with median $SNR\ge 4$ and $1.54\lesssim z_{\rm em}\lesssim 5.16$ to visually analyze narrow $\rm C~IV\lambda\lambda1548,1551$ absorption doublets one by one. A total of 15,999 quasars are found to have at least one appropriate absorption system imprinted on their spectra. From these 15,999 quasar spectra, we have detected 23,336 appropriate $\rm C~IV\lambda\lambda1548,1551$ absorption systems with $W_r\ge0.2$ \AA whose absorption redshifts cover a range of $z_{\rm abs}=1.4544$ --- $4.3704$. The largest values of $W_r$ are 3.19 \AA~ for the $\lambda1548$ absorption line and 2.93 \AA~ for the $\lambda1551$ absorption line, respectively. We find that only a few absorbers show large values of $W_r$. About 1.1\% absorbers of the total absorbers have $W_r\lambda1548\ge2.0$ \AA.
\end{abstract}
\keywords{quasars: absorption lines --- quasars: general}

\section{Introduction}
As we known, galaxies harbor reservoirs of gas know as the circumgalactic medium (CGM). The interactions between galaxies and their surrounding media involve the accretion of material that is required to maintain star formation and the dynamical mechanisms that influence the galactic environment, and move the heavy elements (metal-enriched gas) produced in the stars of galaxies from their places of production into galactic halos, the CGM, and the intergalactic medium (IGM;
Veilleux et al.2005; Fumagalli et al. 2011; Heckman et al. 2011). The latter role would enrich the material for future star formation. The two kinds of interactions are strongly affected by the galactic winds, filament infall, mergers, high velocity clouds, etc. (see Putman et al.2012for a review).

Due to the sensitivity of instruments, however, it is unrealistic to directly observe the fine structures and ingredients of the gaseous clouds harbored by the galaxies, especially for faint galaxies. Fortunately, absorption features are often detected in the spectra of distant objects when their sightlines go through
the foreground (1) high velocity clouds, (2) IGM, (3) filament gas, (4) CGM, etc. These absorption features are a sensitive and unbiased tool for probing the gases of the universe from early periods (Meiksin2009). Absorption features imprinted in the quasar spectra provide us with an excellent opportunity to investigate the gaseous content (e.g., densities, ionization structures, metallicities, temperatures, and kinematics) of extragalactic objects. These investigations do not depend on the luminosity and redshift of the corresponding quasar, or on the velocity and luminosity of the extra-galactic object which
might otherwise be invisible. The destinies of the galaxies are ultimately connected to the fate of gas within them, and therefore studies of quasar absorption lines are helpful to our understanding of the properties of galaxies (e.g., Zibetti et al. 2007;M\'enard et al.2011; Chen 2013; Burchett et al. 2013; Nielsen et al.2013; Kacprzak et al.2013).

Neutral- to high-ionization absorption lines have been observed in quasar spectra (e.g., Misawa et al.2007; Tombesi et al.2011; Chen et al. 2013a, 2013d; Chen \& Qin 2013c; Gupta et al.2013). These absorption features have historically been divided into three classes according to their line widths: (1) broad absorption lines (BALs) with line widths of no less than $2000~km~s^{-1}$ at depths>10\% below the continuum, (2) narrow absorption lines (NALs) with line widths of less than a few hundred $km~s^{-1}$, and (3) mini-BALs with intermediate line widths between those of BALs and NALs (e.g., Weymann et al.1979; Rodr\'iguez Hidalgo et al.2011; Filiz Ak et al.2012; Hamann et al.2013).

Using the quasar spectra of the Sloan Digital Sky Survey (SDSS; York et al.2000), many authors have conducted campaigns to systematically search for narrow metal absorption lines (e.g., Quider et al. 2011; Qin et al. 2013;Zhu \& M\'enard 2013; Cooksey et al. 2013; Seyffertetal.2013). The SDSS program continued with the Third SDSS (SDSS-III) using the same 2.5 m telescope (Gunn et al.2006; Rossetal.2012). This program collected data from 2008 to 2014. The Baryon
Oscillation Spectroscopic Survey (BOSS) is the main dark time legacy survey of SDSS-III (P\^aris et al. 2012),  which has obtained more than 150,000 quasar spectra with $z_{\rm em}>2.15$ during its five year run time. The BOSS spectrograph has a wavelength range of 3600 \AA --- 10400 \AA~ at a resolution of $1300<R<3000$. The SDSS Data Release Nine (SDSS-DR9), which is the first spectral data release of BOSS to the public, includes 87,822 quasars detected over an area of $3275~deg^2$ (P\^a aris et al. 2012).

This paper is the second part in a series aiming to analyze absorption lines in the BOSS quasar spectra. We intend to search for narrow absorption doublets
(e.g., $\rm Mg~II\lambda\lambda2796,2803$, and $\rm C~IV\lambda\lambda1548,1551$) in the BOSS quasar spectra. In the first paper, Chen et al. (2014a, 2014b; hereafter, the two papers are referred to asPaper I), we focused on narrow C IV absorption doublets. Paper I concerned those quasars with $z_{\rm em}\le2.4$
and a high signal-to-noise ratio (S/N) for their spectra. These criteria limit the detection to 15,278 appropriate quasars for the detection of narrow C IV absorption features. Of these 15,278 quasars, 6291 quasars are observed to host C IV absorption system(s). In Paper I, we obtained a catalog of 9417 narrow
C IV systems with $z_{\rm abs}=1.4544$ --- $2.2805$.

This work continues the detection of Paper I by expanding the quasar sample to those quasars with $z_{\rm em}>2.4$ to search for narrow CIV absorption doublets in the BOSS quasar spectra.

We describe our data analysis in Section 2 and present the statistical properties of absorptions in Section 3. The summary is presented in Section 4.

\section{Data analysis}
This work continues the analysis of Paper I by expanding the quasar sample to those quasars with $z_{\rm em}>2.4$. We adopt the same methods to construct quasar samples and detect narrow $\rm C~IV\lambda\lambda1548,1551$ doublets. Here, we simply summarize the main steps involved in the quasar selection and absorption line detection procedures. Readers can refer to Paper I for more details.

In order to avoid the confusions from $\rm Ly\alpha$, $\rm O~I\lambda1302$ and $\rm S~II\lambda1304$ absorptions and the noisy region of the spectra, the spectra regions shortward of 1310 \AA in the rest frame or shortward of 3800 \AA in the observed frame are not considered. In addition, to detect as much of the intervening C iv absorption doublets as possible, we constrain our analysis to the data range within $10,000~km~s^{-1}$ blueward of the quasar system. These
cuts reduce the number of SDSS DR9 quasars from 87,822 to 70,336.

Each quasar spectrum has a value of the median S/N (median S/N). Based on the median value of the median S/N of these 70,336 quasars, median $\rm median~SNR\ge4$ is adopted to alleviate the noise confusion superposed on the spectra with low values of S/N. This further reduces the number of quasar samples from 70,336 to 37,241.

As a continuation of the work of Paper I, here we only concern ourselves with high-redshift quasars with $z_{\rm em}>2.4$. Considering all of the above limitations, we obtain 21,963 appropriate quasars. Figure 1 shows the limits of the median S/N and the redshift. The emission redshift distribution of
21,963 quasars is shown by the red line in Figure 2.

\begin{figure}
\vspace{3ex} \centering
\includegraphics[width=7 cm,height=6 cm]{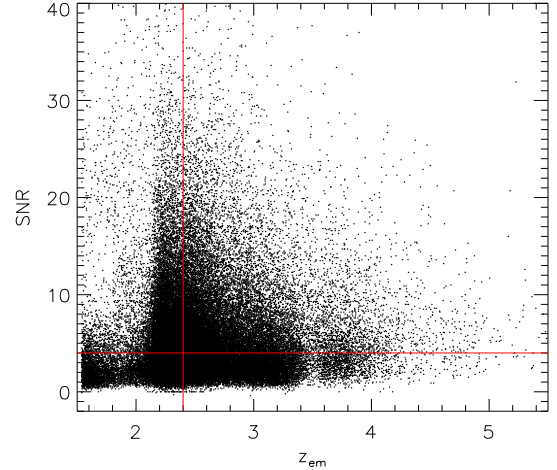}\vspace{3ex}
\caption{Plot of the median S/N vs. the emission redshift. The horizontal red line stands for $S/N=4$ and the vertical red line stands for $z_{em}=2.4$, which are used to construct quasar samples. Paper I contained quasars with $SNR\ge4$ and $z_{em}\le2.4$. In this paper, we detect Civ absorption on the
spectra of the quasars with $SNR\ge4$ and $z_{em}>2.4$.}
\end{figure}

For each quasar, a combination of the cubic spline function and Gaussian functions was invoked to derive a pseudo-continuum in an iterative fashion (e.g., Nestor et al.2005; Chen et al.2013a,2013d,2013b; Chen \& Qin2013c). Civ absorption doublets were detected on the pseudo-continuum fitting normalized spectra. There are several steps in this detection. First, we label the $2\sigma$ flux uncertainty level and get rid of the absorption troughs above this level. Second, BALs are disregarded by the program searching for candidate absorptions. Third, in this program, a Gaussian function is adopted to fit each absorption feature, while the absorption figures with FWHM greater than $800~\rm km~s^{-1}$ are ignored. Fourth, measurements of the equivalent widths at rest-frame ($W_r$) of the candidate absorptions are based on the Gaussian fittings, and estimations of their flux uncertainties are performed via
\begin{equation}
(1+z)\sigma_w=\frac{\sqrt{\sum_i
P^2(\lambda_i-\lambda_0)\sigma^2_{f_i}}}{\sum_i
P^2(\lambda_i-\lambda_0)}\Delta\lambda,
\end{equation}
where, as a function of pixel, $P(\lambda_i-\lambda_0)$, $\lambda_i$ and $\sigma_{f_i}$ are the line profile centered at $\lambda_0$, the wavelength, and the
normalized flux uncertainty, respectively. The signal-to-noise ratio of candidate absorption lines is estimated using the same method adopted by Qin et al. (2013). The value of $\rm 1\sigma$ noise is computed via:
\begin{equation}
\sigma_N=\sqrt{\frac{\sum\limits_{i=1}^M[\frac{F^i_{noise}}{F^i_{cont}}]^2}{M}},
\end{equation}
where $\rm F_{noise}$, $\rm F_{cont}$, and $i$ are the flux uncertainty, the flux of the pseudo-continuum fitting, the pixel in the wavelength range of
1548{\AA}$\rm\times(1+z_{abs})-5${\AA}$\rm<\lambda_{obs}<1551${\AA}$\rm\times(1+z_{abs})+5${\AA}, respectively. Based on the value determined by formula (2), we can obtain the signal-to-noise ratio of the candidate absorption line via
\begin{equation}
SNR^\lambda=\frac{1-S_{abs}}{\sigma_N},
\end{equation}
where $\rm S_{abs}$ is the largest depth within an absorption trough with respect to the pseudo-continuum fitting in the normalized spectra. Fifthly, all absorption features with $W_r>0.2$ \AA~ and $SNR^{\lambda} \ge 2.0$ for both lines of the doublet are kept in the catalog.

Using the same method as in Paper I, we find that 9708 quasars from the quasar sample (containing 21,963 quasars) host narrow Ci vabsorption system(s). From these 9708 quasar spectra, we obtain 13,919 potential intervening Civ absorption systems. The absorption data are provided in Table 1.

\begin{table*}[htbp]
\caption{Catalog of $\rm C~IV\lambda\lambda1548,1551$ absorption
systems} \tabcolsep 1.1mm \centering %\tiny
 \begin{tabular}{cccccccccccccc}
 \hline\hline\noalign{\smallskip}
SDSS NAME & PLATEID & MJD & FIBERID & $z_{\rm em}$ & $z_{\rm abs}$ &
$\rm W_r\lambda1548$ &$N_{\sigma\lambda1548}$& $\rm W_r\lambda1551$&$N_{\sigma\lambda1551}$&$SNR^{\lambda1548}$&$SNR^{\lambda1551}$&$\beta$\\
\hline\noalign{\smallskip}
000014.07+012951.5  &   4296    &   55499   &   0370    &   3.2284  &   2.6728  &   0.70    &   4.38    &   0.68    &   4.00    &   4.2     &   3.8     &   0.13994     \\
000015.17+004833.2  &   4216    &   55477   &   0718    &   3.0277  &   2.5222  &   1.11    &   6.94    &   0.84    &   7.64    &   6.4     &   7.1     &   0.13331     \\
000027.31+013126.1  &   4296    &   55499   &   0382    &   2.5892  &   2.3169  &   0.35    &   4.38    &   0.37    &   4.63    &   3.9     &   4.1     &   0.07874     \\
000041.87-001207.2  &   4216    &   55477   &   0274    &   3.0514  &   2.6206  &   0.37    &   6.17    &   0.42    &   5.25    &   6.0     &   4.9     &   0.11195     \\
000041.87-001207.2  &   4216    &   55477   &   0274    &   3.0514  &   2.6600  &   0.83    &   5.93    &   0.30    &   3.33    &   5.6     &   3.2     &   0.10125     \\
000046.42+011420.8  &   4216    &   55477   &   0742    &   3.7588  &   3.1522  &   1.50    &   8.33    &   1.39    &   8.69    &   7.9     &   8.2     &   0.13552     \\
000046.42+011420.8  &   4216    &   55477   &   0742    &   3.7588  &   3.1777  &   1.39    &   9.93    &   1.09    &   8.38    &   9.4     &   7.7     &   0.12950     \\
000051.56+001202.5  &   4216    &   55477   &   0778    &   3.8953  &   3.3315  &   1.18    &   4.54    &   1.19    &   4.76    &   4.3     &   4.5     &   0.12175     \\
000051.56+001202.5  &   4216    &   55477   &   0778    &   3.8953  &   3.3721  &   2.43    &   9.00    &   2.87    &   10.63   &   8.5     &   10.0    &   0.11255     \\
000057.58+010658.6  &   4216    &   55477   &   0750    &   2.5493  &   2.0827  &   0.48    &   4.36    &   0.80    &   8.00    &   4.1     &   7.7     &   0.14002     \\
\hline\hline\noalign{\smallskip}
\end{tabular}
\\
 \footnote[]~Note---$N_{\sigma}=\frac{W_r}{\sigma_{W}}$ represents the significant level of the detection. $\beta=v/c=((1+z_{em})^2-(1+z_{abs})^2)/((1+z_{em})^2+(1+z_{abs})^2)$.
\end{table*}

\section{Statistical properties of the absorbers}
\subsection{Properties of the absorbers in catalog (II)}
This paper is the second in a series on the analysis of absorption lines in the quasar spectra of BOSS, which expands the quasar sample to quasars with $z_{\rm em}>2.4$. 21,963 appropriate quasars are adopted to search for narrow $\rm C~IV\lambda\lambda1548,1551$ absorption doublets. Among these, 9708 quasars are observed to host appropriate C iv absorption feature imprinted on their spectra. The distributions of the emission redshifts of the 21,963 quasars and the 9708 quasars are shown in Figure2. In the 9708 quasar spectra, we detect 13,919 narrow $\rm C~IV\lambda\lambda1548,1551$ absorption doublets with $1.8784\le z_{\rm abs}\le4.3704$ , and the distribution of these absorption redshifts are also shown in Figure 2. (Note that these absorption doublets only refer  to catalog II, and those contained in Paper I only refer to catalog I.)

\begin{figure}
\vspace{3ex}\centering
\includegraphics[width=7 cm,height=6 cm]{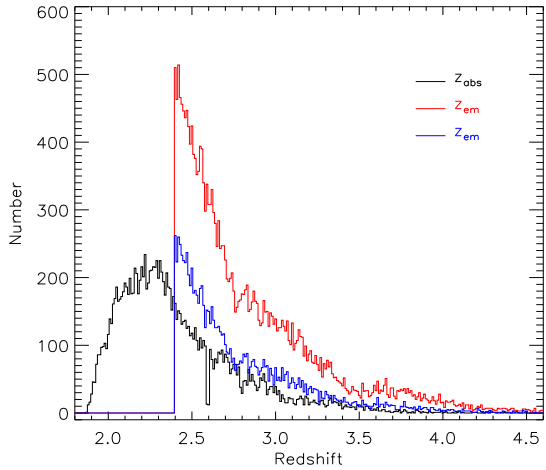}\vspace{3ex}
\caption{Distributions of redshifts for catalog II. The red line represents the distribution of the emission redshift of the 21,963 quasars used to detect Civ absorption systems, and the blue line is for that of the 9708 quasars detected with Civ absorptions. The black line represents the distribution of the redshift of the Civ absorption doublets.}
\end{figure}

In Figure 3, we plot the total redshift path covered by the absorbers of catalog II as a function of the S/N of the absorption features, which is calculated by 
\begin{equation}
Z(SNR^{\lambda1548})=\sum_{i=1}^{N_{spec}}\int_{z_i^{min}}^{z_i^{max}}g_i(SNR^{\lambda1548},z)dz,
\end{equation}
where $\rm z_i^{max}$ and $\rm z_i^{min}$ are the redshifts corresponding to the maximum and minimum wavelengths in the survey spectral region of quasar $i$, respectively; $\rm g_i(SNR^{\lambda1548},z)=1$ if $\rm SNR^{lim}~\le SNR^{\lambda1548}$, otherwise $\rm g_i(SNR^{\lambda1548},z)=0$ (see also Qin et al. 2013; Paper 1).

\begin{figure}
\centering
\includegraphics[width=7cm,height=6cm]{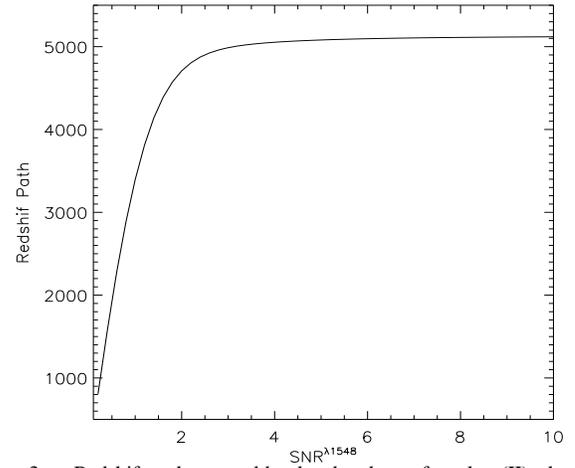}
\caption{Redshift path covered by the absorbers of catalog (II), shown as a function of $\rm SNR^{\lambda1548}$.}
\end{figure}

In Figure 4, we show the Wr distributions of detected $\rm C~IV\lambda\lambda1548,1551$ absorption doublets in catalog II, which are similar to those in catalog I. Smooth tails out to $W_r\approx3.0$ \AA~ can clearly be seen in these distributions. The largest and median values of $W_r\lambda1548$ are 2.95 \AA~ and 0.60 \AA~ respectively, and those of $W_r\lambda1551$ are 2.93 \AA~ and 0.47 \AA~ respectively. In catalog II, only a few absorbers with large $W_r$ are detected, with only 1.0\% (137/13919) and 16.9\% (2349/13919) absorbers of the total having $W_r\lambda1548\ge2.0$ \AA~ and $1.0$ \AA$\le W_r\lambda1548<2.0$ \AA, respectively. Most absorbers show small or medium values of absorption strengths, with about 47.2\% (6564/13919) and 35.0\% (4869/13919) absorbers of the total having $0.5$ \AA$\le W_r\lambda1548<1.0$ \AA~ and $0.2$ \AA$\le W_r\lambda1548<0.5$ \AA,~ respectively. These proportions are similar to those in Paper I.

\begin{figure}
\vspace{3ex}\centering
\includegraphics[width=7 cm,height=6 cm]{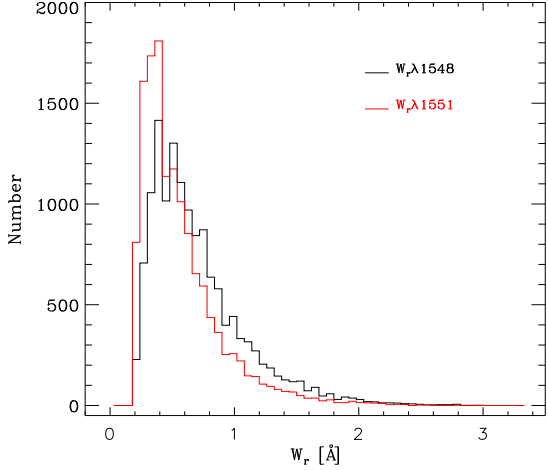}\vspace{3ex}
\caption{Distributions of the $\rm W_r$ of the $\rm C~IV$ absorption lines detected in catalog (II). The black line stands for the $\rm \lambda1548$ absorptions, and the red line stands for the $\rm \lambda1551$ absorptions.}
\end{figure}

The saturated degree of absorptions can be evaluated by the $W_r$ ratio ($DR$; Str\"omgren 1948). Theoretical values of $DR$ of the $\rm C~IV\lambda\lambda1548,1551$ resonant doublet can vary from $1.0$ for completely saturated absorption to 2.0 for completely unsaturated absorption (e.g., Sargent et al. 1988). In Figure 5, we show the $DR$ of the $\rm C~IV\lambda\lambda1548,1551$ resonant doublet ($W_r\lambda1548/W_r\lambda1551$) in catalog (II) where the theoretical boundaries of the completely unsaturated absorption and completely saturated absorption are plotted with red dash lines. A maximum value (4.7) of the $DR$ and a minimum value (0.2) can be seen from the Figure. There are about 20.4\% (2844/13919) and 7.1\% (993/13919) absorbers with $DR<1.0$ and $DR>2.0$, respectively. Meanwhile, there are about 72.4\% (10082/13919) absorbers with $1.0\le DR \le 2.0$. These proportions are similar to those in Paper I.
Some narrow absorption features cannot be resolved by a low-/middle-resolution spectrograph, which may result in blending of narrow absorption features. BOSS spectra have a resolution of $1300<R<3000$. These low-/middle-resolution spectra can lead to some very narrow C iv absorptions suffering from blending with unrelated absorption features. The ``false positive" lines, which happen to have the same separation as the two lines of the C iv doublet but essentially no relation to each other, may confuse the identification of the C iv absorption doublet. We guess that line blending and this kind of ``false positive" line would be the main reason accounting for the fact that about a quarter of the narrow Civ systems lie outside the theoretical limits of the $W_r$ ratio. This conjecture could be checked by Monte Carlo simulations. That is an interesting issue that deserves a detailed investigation, and we intend to do so in a later paper.

\begin{figure}
\vspace{3ex}\centering
\includegraphics[width=7.5 cm,height=6.5 cm]{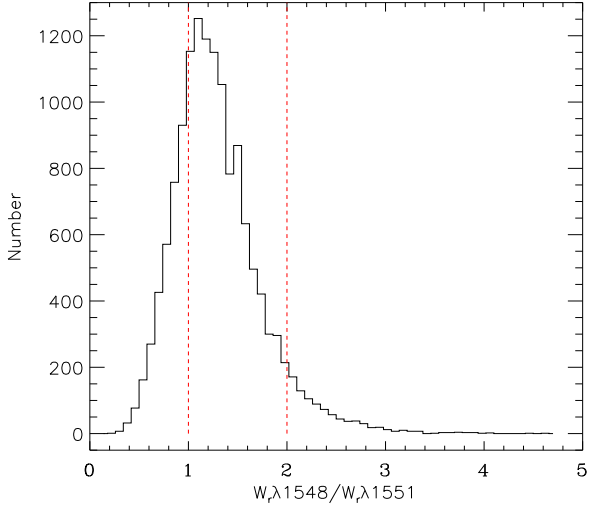}\vspace{3ex}
\caption{Distribution of the $W_r$ ratio of the $\rm C~IV$ doublets in catalog (II). The red dash lines are the theoretical boundaries of completely saturated ($\rm W_r\lambda1548/W_r\lambda1551 = 1.0$) and unsaturated ($\rm W_r\lambda1548/W_r\lambda1551 = 2.0$) absorptions, respectively.}
\end{figure}

We calculate the frequency of detected $\rm C~IV$ absorption doublets ($f_{\rm NALs}$) to give an expression to the false positives/negatives of $\rm C~IV$ absorption doublets as follow
\begin{equation}
f_{NALs}=\lim_{\bigtriangleup SNR\rightarrow0}\frac{\bigtriangleup N_{abs}}{\bigtriangleup N_{sdp}}
\end{equation}
where $\bigtriangleup N_{\rm abs}$ is the number of detected $\rm C~IV$ absorption doublets, and $\bigtriangleup N_{\rm sdp}$ is the number of data points in signal-to-noise ratio bin $\bigtriangleup SNR$. The deriving $f_{\rm NALs}$ is shown in Figure 6. It clearly shows a smooth distribution of the $f_{\rm NALs}$ in the range of $SNR^{\lambda1548}\gtrsim4$, which implies a completed detection when the signal-to-noise ratio is larger than 4. This is similar to that in Paper 1.

\begin{figure}
\centering
\includegraphics[width=7cm,height=6cm]{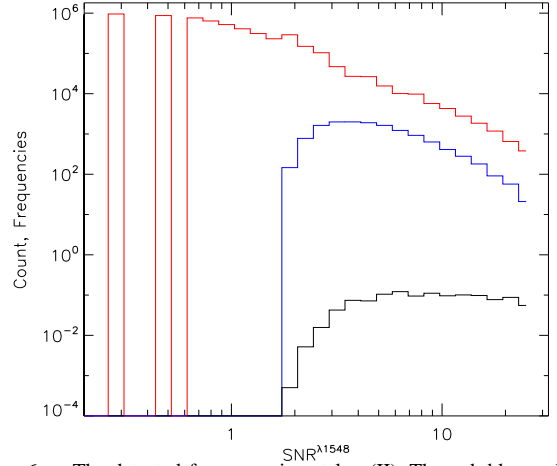}
\caption{The detected frequency in catalog (II). The red, blue and black lines represent the number of data points, number of detected $\rm C~IV$ absorptions, and frequency of NALs in catalog (II), respectively.}
\end{figure}

Figure 6 also suggests a significant incomplete detection when $SNR^{\lambda1548}\lesssim4$, whose missing rate ($f_{MR}$) can be evaluated in several bins of the signal-to-noise ratio by
\begin{equation}
f_{MR}=\frac{\overline{f_{NALs}}-f_{NALs}}{\overline{f_{NALs}}},
\end{equation}
where $\overline{f_{NALs}}$ and $f_{NALs}$ are the average frequency of NALs in the range of $SNR^{\lambda1548}>4$, and the frequency of NALs in the corresponding signal-to-noise ratio bin, respectively. The results are provided in Table 2.

\begin{table}
\caption{The missing rate of absorption systems with $SNR^{\lambda1548}\le 4$} \tabcolsep 2mm \centering %\tiny
 \begin{tabular}{ccccccc}
 \hline\hline\noalign{\smallskip}
SNR bin & [2.0,2.5] & [2.5,3.0] & [3.0,3.5] & [3.5,4.0]\\
\hline\noalign{\smallskip}
$f_{MR}$&0.88&0.56&0.53&0.07\\
\hline\hline\noalign{\smallskip}
\end{tabular}
\end{table}

\subsection{Properties of the absorbers in catalogs (I - II)}
As the first two papers in a series concerned with searching absorption lines in the BOSS quasar spectra, we have collected a sample of 37,241 appropriate quasars with median $\rm SNR\ge4$ and $1.54\lesssim z_{\rm em}\lesssim 5.16$ to detect potential intervening $\rm C~IV\lambda\lambda1548,1551$ absorptions. 
From the quasar sample, we have found 15,999 quasars with at least one appropriate $\rm C~IV\lambda\lambda1548,1551$ absorption doublets imprinted on their spectra. From these 15,999 quasar spectra, we have detected 23,336 narrow $\rm C~IV\lambda\lambda1548,1551$ absorption systems with $z_{\rm abs}=1.4544$ --- $4.3704$. Shown In Figure 7, we display distributions of emission redshifts of the 37,241 quasars and the 15,999 quasars, as well as the absorption redshifts of the 23,336 absorbers.

\begin{figure}
\vspace{3ex}\centering
\includegraphics[width=7 cm,height=6 cm]{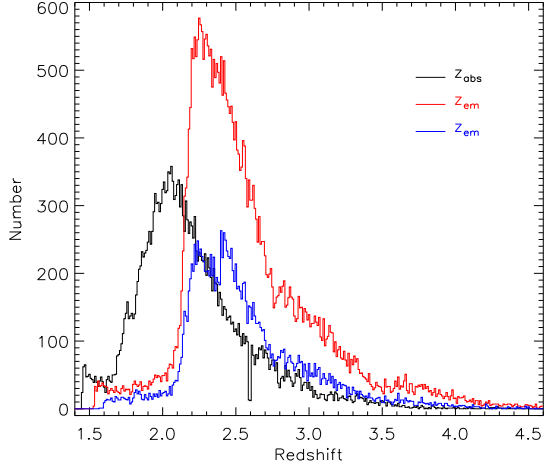}\vspace{3ex}
\caption{Distributions of redshifts included in catalogs (I - II). See Figure 2 for the meanings of the lines.}
\end{figure}

The redshift path of catalogs I and II is calculated using formula (4) as well. The result is plotted in Figure8where the redshift paths covered by catalogs I and II are also shown with dot-dashed and dashed lines, respectively.

\begin{figure}
\centering
\includegraphics[width=7cm,height=6cm]{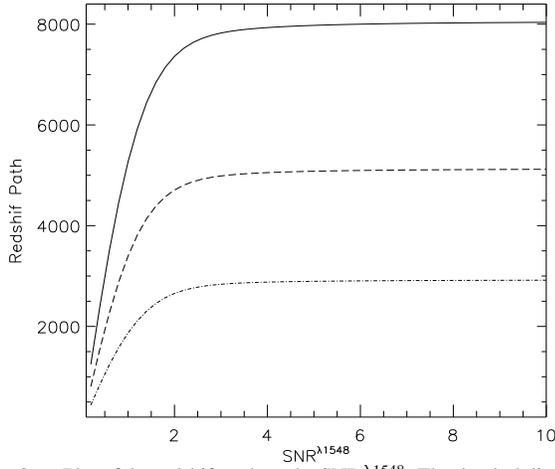}
\caption{Plot of the redshift path vs the $\rm SNR^{\lambda1548}$. The dot dash line is for that of catalog (I); the dash line is for that of catalog (II); and the solid line is the sum of that of these two catalogs.}
\end{figure}

The $W_r$ distributions are plotted in Figure 9, which show tails up to $W_r\approx3$ \AA.~ The largest and median absorption strengths of $\lambda1548$ lines are $\rm W_r\lambda1548= 3.19$ \AA~ and 0.61 \AA, respectively, and those of $\lambda1551$ are $\rm W_r\lambda1551= 2.93$ \AA~ and 0.48 \AA, respectively. We find that, in catalogs (I - II), there are a few absorbers with large $W_r$, namely, only 1.1\% (249/23336) and 17.8\% (4124/23336) absorbers of the total with $W_r\lambda1548\ge2.0$ \AA~ and $1.0$ \AA$\le W_r\lambda1548<2.0$ \AA, respectively. Most of the absorbers show small or middle values of absorption strengths. That is, there are about 46.6\% (10879/23336) and 34.6\% (8084/23336) absorbers of the total with $0.5$ \AA$\le W_r\lambda1548<1.0$ \AA~ and $0.2$ \AA$\le W_r\lambda1548<0.5$ \AA, respectively.

\begin{figure}
\vspace{3ex}\centering
\includegraphics[width=7 cm,height=6 cm]{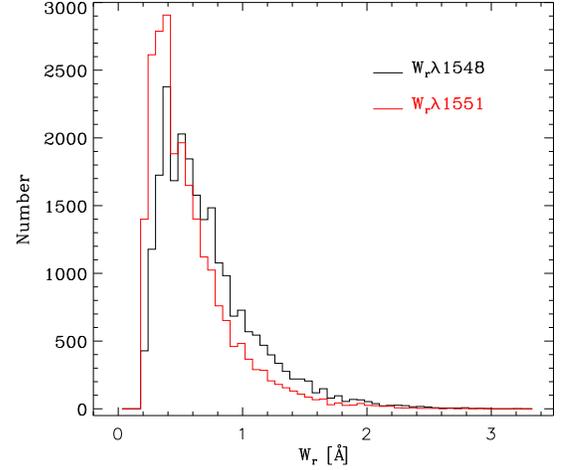}\vspace{3ex}
\caption{Distributions of $\rm W_r$ of $\rm C~IV$ absorptions that are included in catalogs (I - II). The black and red lines represent the $\rm \lambda1548$ and $\rm \lambda1551$ absorptions, respectively.}
\end{figure}

The distribution of the $W_r$ ratio is displayed in Figure 10. The $W_r$ ratio has a maximum value of 4.7, and has a minimum value of 0.2. There are about 21.2\% (4958/23336) and 6.7\% (1565/23336) absorbers with $DR<1.0$ and $DR>2.0$ respectively. And there are about 72.1\% (16813/23336) absorbers with $1.0\le DR \le 2.0$.

\begin{figure}
\vspace{3ex}\centering
\includegraphics[width=7.5 cm,height=6.5 cm]{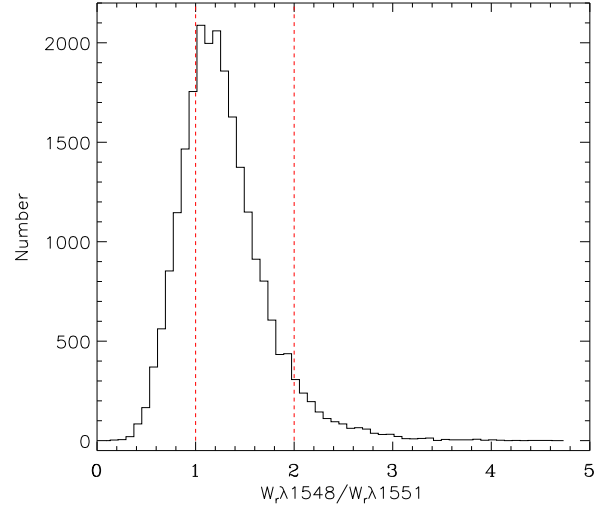}\vspace{3ex}
\caption{Distribution of the $\rm W_r$ ratios of the $\rm C~IV$ doublets in catalogs (I - II) (the solid line). See Figure 5 for the meanings of other lines.}
\end{figure}

The detected frequency of $\rm C~IV$ absorptions is calculated using formula (5). The deriving result is plotted in Figure 11, where the distributions of $f_{\rm NALs}$ obtained from catalogs (I) and (II) are displayed with dot dash and dash lines, respectively. Smooth curves are the distributions of $f_{\rm NALs}$ with $SNR^{\lambda1548}\gtrsim4$.

\begin{figure}
\centering
\includegraphics[width=7cm,height=6cm]{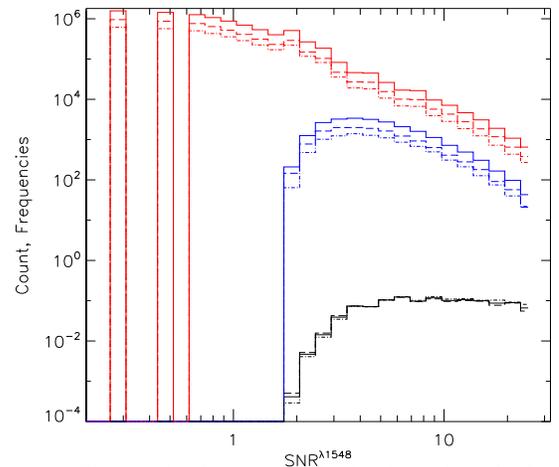}
\caption{The detection frequency vs the signal-to-noise ratio. See Figure 6 for meanings of color lines. The dot dash line is for those of catalog (I); the dash line is for those of catalog (II); and the solid line is for those of the sum of the two catalogs.}
\end{figure}

We estimate the missing rate of absorption systems with $SNR^{\lambda1548}\le 4$ in several bins of the signal-to-noise ratio using formula (6). The deriving results are provided in Table 3.

\begin{table}
\caption{The missing rate of absorption systems with $SNR^{\lambda1548}\le 4$} \tabcolsep 2mm \centering
 \begin{tabular}{ccccccc}
 \hline\hline\noalign{\smallskip}
SNR bin & [2.0,2.5] & [2.5,3.0] & [3.0,3.5] & [3.5,4.0]\\
\hline\noalign{\smallskip}
$f_{MR}$&0.91&0.68&0.62&0.20\\
\hline\hline\noalign{\smallskip}
\end{tabular}
\end{table}

\section{Summary}
This paper continues the work of Paper I aiming to detect absorption lines by expanding the quasar sample to those quasars with $z_{\rm em}>2.4$. This sample contains 21,963 quasars, of which 9708 quasars have at least one potential intervening $\rm C~IV\lambda\lambda1548,1551$ absorption systems with $W_r\ge0.2$ \AA~ for both lines. From these quasar spectra, we have detected 13,919 appropriate $\rm C~IV\lambda\lambda1548,1551$ absorption systems with $z_{\rm abs}=1.8784$ --- $4.3704$. About 35.0\%, 47.2\%, 16.9\%, and 1.0\% absorbers of the total have $W_r\lambda1548=0.2$ --- $0.5$ \AA,~ $W_r\lambda1548=0.5$ --- $1.0$ \AA,~ $W_r\lambda1548=1.0$ --- $2.0$ \AA,~ and $W_r\lambda1548\ge2.0$ \AA,~ respectively.

The absorption doublets detected in this paper refer to catalog II, and those from Paper I refer to catalog I. Catalogs I and II contain a total of 15,999 quasars that have been detected as hosting appropriate Civ absorptions imprinted on their spectra. These quasars are selected from a sample of 70,336 quasars in the limit of median $SNR\ge4$. From these quasar spectra, we have detected 23,336 narrow Civ absorption systems with $z_{\rm abs}=1.4544$ --- $4.3704$. The largest absorption strengths are 3.19 \AA and 2.93 \AA for the $\lambda1548$ and  $\lambda1551$ lines, respectively. Only a few absorbers show large values of $W_r$, with most of the absorbers having small or middle values of $W_r$. There are about 34.6\%, 46.6\%, 17.8\%, and 1.1\% absorbers of the total with $W_r\lambda1548=0.2$ --- $0.5$ \AA, $W_r\lambda1548=0.5$ --- $1.0$ \AA,~ $W_r\lambda1548=1.0$ --- $2.0$ \AA,~ and $W_r\lambda1548\ge2.0$ \AA,~ respectively.

\acknowledgements  We thank the anonymous referee for helpful comments and suggestions. This work was supported by the National Natural Science Foundation of China (NO. 11363001), the Guangxi Natural Science Foundation (2012jjAA10090), the Guangzhou technological project (No. 11C62010685), Guangdong Province
Universities and Colleges Pearl River Scholar Funded Scheme(GDUPS)(2009), Yangcheng Scholar Funded Scheme(10A027S), and the Guangxi university of science and technology research projects (NO. 2013LX155).

Funding for SDSS-III has been provided by the Alfred P. Sloan Foundation, the Participating Institutions, the National Science Foundation, and the U.S. Department of Energy Office of Science. The SDSS-III web site is http://www.sdss3.org/.

SDSS-III is managed by the Astrophysical Research Consortium for the Participating Institutions of the SDSS-III Collaboration including the University of Arizona, the Brazilian Participation Group, Brookhaven National Laboratory, Carnegie Mellon University, University of Florida, the French Participation Group, the German Participation Group, Harvard University, the Instituto de Astrofisica de Canarias, the Michigan State/Notre Dame/JINA Participation Group, Johns Hopkins University, Lawrence Berkeley National Laboratory, Max Planck Institute for Astrophysics, Max Planck Institute for Extraterrestrial Physics, New Mexico State University, New York University, Ohio State University, Pennsylvania State University, University of Portsmouth, Princeton University, the Spanish Participation Group, University of Tokyo, University of Utah, Vanderbilt University, University of Virginia, University of Washington, and Yale University.

\end{document}